\documentclass[twocolumn,superscriptaddress,amssymb,amsmath,nobibnotes,aps,prd,showpacs,nofootinbib]{revtex4-1}
\pdfoutput=1

\usepackage{graphicx,subfigure,bm,color,psfrag,hyperref}
\usepackage{amsfonts}
\usepackage{lipsum}
\usepackage{caption}
\usepackage{mathtools}

\newcommand{\be}{\begin{equation}}
\newcommand{\ee}{\end{equation}}

\begin{document}

\title{Cosmological implications of scale-independent energy-momentum squared gravity: \\ Pseudo nonminimal interactions in dark matter and relativistic relics}

\author{\"{O}zg\"{u}r Akarsu}
\email{akarsuo@itu.edu.tr}

\author{Nihan Kat{\i}rc{\i}}
\email{nihan.katirci@itu.edu.tr}
\affiliation{Department of Physics, \. Istanbul Technical University, Maslak 34469 \. Istanbul, Turkey}

\author{Suresh Kumar}
\email{suresh.kumar@pilani.bits-pilani.ac.in}
\affiliation{Department of Mathematics, BITS Pilani, Pilani Campus, Rajasthan-333031, India}

\author{Rafael C. Nunes}
\email{rafadcnunes@gmail.com}
\affiliation{Departamento de F\'isica, Universidade Federal de Juiz de Fora,
36036-330, Juiz de Fora, Minas Gerais, Brazil}

\author{M. Sami}
\email{samijamia@gmail.com}
\affiliation{Centre for Theoretical Physics, Jamia Millia Islamia, New Delhi, India}

\begin{abstract}
In this paper, we introduce a scale-independent energy-momentum squared gravity (EMSG) that allows different gravitational couplings for different types of sources, which may lead to scenarios with many interesting applications/implications in cosmology. In the present study, to begin with, we study a modification of the $\Lambda$ cold dark matter ($\Lambda$CDM) model, where photons and baryons couple to the spacetime as in general relativity, while the cold dark matter and relativistic relics (neutrinos and any other relativistic relics) couple to the spacetime in accordance with EMSG. This scenario induces pseudo nonminimal interactions on these components, leading to modification at both the background and perturbative levels. A consequence of this scenario is that the dimensionless free parameter of the theory may induce direct changes on the effective number of the relativistic species, without the need to introduce new extra species. In order to quantify the observational consequences of the cosmological scenario, we use the cosmic microwave background Planck data (temperature, polarization, and lensing power spectrum) and baryonic acoustic oscillations data. 
We find that the free model parameter is too small to induce statistically significant corrections on the $\Lambda$CDM model due to EMSG. We deduce that the model presented here is quite rich with promising cosmological applications/implications that deserve further investigations.
\end{abstract}

\maketitle
\section{Introduction}
\label{sec:intro}
Being simple and in reasonably good agreement with the currently available high-precision data \cite{Komatsu:2010fb,Aubourg:2014yra,Planck2015}, the minimal $\Lambda$ cold dark matter ($\Lambda$CDM) model is, so far, the most successful cosmological model describing the dynamics as well as the large-scale structure of the observable Universe. However, it suffers from theoretical inconsistencies relevant to the cosmological constant \cite{Weinberg:1988cp,Peebles:2002gy,Padmanabhan:2002ji}, and tensions between the observational constraints obtained from different data sets: e.g., data from some local observations do not agree with the predictions of the minimal $\Lambda$CDM model \cite{tension01, tension02, tension03, Aubourg:2014yra,Zhao:2017cud,Bullock:2017xww,Freedman:2017yms}.
Moreover, the observations suggesting small deviations from $\Lambda$CDM require profound modifications in fundamental theories. We do not have a promising and concrete fundamental theory giving rise to dark energy (DE) models (see Refs. \cite{Copeland:2006wr,Caldwell:2009ix} for a review) more general than the cosmological constant $\Lambda$ that would account for these small deviations. The situation does not seem to improve in the broad avenue followed by many cosmological studies introducing modifications to general theory of relativity (GR) by generalizing the gravitational Lagrangian away from the linear function of scalar curvature $R$ (see Refs. \cite{Clifton:2011jh,DeFelice:2010aj,Capozziello:2011et,Nojiri:2017ncd,Nojiri:2010wj} for a review). Alternatively, we can look for the small deviations from the $\Lambda$CDM model by keeping $\Lambda$ to account for the accelerated expansion as it is but introducing modifications relevant to the dark sector, viz., cold dark matter (CDM) and relativistic relics (neutrinos and any other relativistic relics), of the model by (i) manipulating physical properties of dark sector sources, (ii) introducing extra species in the dark (hidden) sector that may not easily be distinguished from the usual ones, and (iii) introducing a modified gravity theory that manipulates the gravitational coupling of the dark sector sources only. The latter approach that we follow in this paper, to our knowledge, is not common in the literature as much as the former two, since presumably it implies violation of the equivalence principle (EP) underlying the Einstein's general theory of relativity (for studies considering such an approach, see, for instance, Refs. \cite{Hui:2009kc,peebles10,Mohapi:2015gua}).  

The dark sector of our Universe may be more featured than the conventional picture of weakly interacting CDM particles and neutrinos. For instance, the possibility of nonminimal interactions of dark matter have been intensely investigated in the literature from different motivations and perspectives, e.g., dark matter interaction with DE \cite{DM_DE01,DM_DE02,DM_DE03}, neutrinos \cite{DM_nu01, DM_nu02}, baryons \cite{DM_B01,DM_B02,DM_B03}, photons \cite{DM_gamma01,DM_gamma02} and dark radiation \cite{DM_DR01,DM_DR02}. In this paper, we follow a novel approach assuming that CDM and relativistic relics (neutrinos and any other relativistic relics) are interacting only gravitationally and have the conventional intrinsic properties, namely, have the equations of state as usual as $p=0$ and $p/\rho=1/3$, respectively, but couple to the spacetime in accordance with a modified gravity, namely, the \textit{scale-independent energy-momentum squared gravity}, which allows different gravitational couplings for the different species of the sources, and thereby induces a pseudo nonminimal interaction of each species culminating in modifications at the background and perturbative levels.

Gravity theories violating EP can naturally result from generalizing the form of the matter Lagrangian, $\mathcal{L}_{\rm m}$, in a nonlinear way, for instance, to some analytic function of the scalar $T^{2}=T_{\mu \nu }T^{\mu \nu }$ formed from the energy-momentum tensor (EMT), $T_{\mu \nu }$, of the material stresses \cite{Katirci:2014sti}, rather than generalizing the gravitational Lagrangian away from the linear function of scalar curvature, $R$. Such generalizations of GR include new types of contributions of the material stress to the right-hand side of the Einstein's field equations, without invoking some new types of sources (for other similar types of theories, see, e.g. Refs. \cite{Harko:2010mv,Harko:2011kv}). One may look for a modification on the left-hand side of the Einstein's field equations corresponding to these new types of contributions of the material stress, but this might not be trivial or even possible \footnote{It is well known that modifications to the matter sector of a gravity theory can sometimes be recast as a modification to its gravitational sector, though, even if it could be done, namely, these two gravity theories would lead mathematically to the same equations of motion, this would not imply that these two gravity theories are physically equivalent. The investigation of EMSG in this regard seems nontrivial, in particular, due to the violation of the EP as well as of the local energy conservation, which actually underly the features of the scale-independent EMSG that we consider in this work. Indeed, the aspects of EMSG such as in what type of gravity theories and under what conditions the new contributions of the material stress to the right-hand side of the Einstein's field equations due to the EMSG could be recast as a modification to the gravitational sector of the considered gravity theory, are open questions for the researchers interested in EMSG. } (see Refs. \cite{Akarsu:2017ohj,Board:2017ign,Akarsu:2018zxl} for some further relevant discussions), and thereby such types of modified gravity theories are promising in constructing novel cosmological models. A particular example of this type of generalizations in the form $f(T^{2})=\alpha T^{2}$, dubbed as ``energy-momentum squared gravity"  (EMSG), is studied in Refs. \cite{Roshan:2016mbt,Akarsu:2018zxl,Nari:2018aqs}, and a more general one, in the form $f(T^{2})=\alpha(T^{2})^{\eta}$, dubbed as \textit{energy-momentum powered gravity} (EMPG) is studied in Refs. \cite{Akarsu:2017ohj,Board:2017ign}. The higher-order matter terms in EMSG here are reminiscent of the terms (corrections) that arise naturally in loop quantum gravity \cite{Ashtekar:2006wn,Ashtekar:2011ni}, and those in the brane world models \cite{Brax:2003fv}. In the EMPG model, $(T^{2})^{\eta}$ modification becomes effective at high energy densities (e.g., in the early Universe \cite{Board:2017ign,Akarsu:2018zxl}, namely, the initial singularity can be replaced with an initial bounce, and domination of spatial anisotropy about the initial singularity can be avoided) for the cases $\eta>1/2$ and at low energy densities (e.g., in the dynamics of the late Universe, namely, the case $\eta=0$ leads to the $\Lambda$CDM-type model, and $\eta \sim 0$ leads to a $w$CDM-type model, though the Universe is sourced by pressureless matter only) for the cases $\eta<1/2$. See Refs. \cite{Akarsu:2017ohj} for more details.

In this paper, we propose a modified theory of gravitation constructed by the addition of the term $f\left(T_{\mu\nu}T^{\mu\nu}, \mathcal{L}_{\rm m}\right)$ to the Einstein-Hilbert (EH) action with a cosmological constant and investigate a particular case, $f(T^{2})=\alpha (T^{2})^{1/2}$ (the case $\eta=1/2$ of EMPG). It is particular in the sense that the contributions to Einstein's field equations due to EMSG come with the same power as the usual terms from the EH part of the action. So, the EMSG modification would affect the field equations \emph{independent of the energy density scale} considered, and thereby, this case may be dubbed as a \textit{scale-independent EMSG}. This theory is at once able to provide many features that researchers have been considering separately: (i) Sources with different equations of state couple to the spacetime differently \footnote{It is possible to see such features in modified gravity theories including new type of contributions of the material stress to the right-hand side of the Einstein's field equations like in EMSG (e.g., \cite{Harko:2010mv,Harko:2011kv,Harko:2014aja}), as well as in scalar-tensor theories of gravity, e.g., in Brans-Dicke theory in the Einstein frame since different stresses have different traces in $T_{\rm m}\phi$ coupling (see Refs. \cite{Hui:2009kc} for details.).}. (ii) Because the modification in the Lagrangian is done by an analytical function of the EMT, it is possible to define different coupling constants for the EMTs of different sources \footnote{The first example considering such a feature could be seen in \cite{Damour90} (motivated by Ref. \cite{Wetterich:1994bg} based on string theory) where it is done by constructing a generalized Brans-Dicke theory in which the scalar field couples with different strengths to baryons and to CDM (see, e.g., Refs. \cite{amendola2000,amendola2002} following the similar idea and Refs. \cite{Farrar:2003uw,GarciaBellido:1992de} summarizing the history of the idea). We also would like to note here that $f(R,L_{\rm m})$, $f(R,T)$, $f(\mathcal{T},T)$ theories \cite{Harko:2010mv,Harko:2011kv,Harko:2014aja}  also include new type of contributions of the material stress to the right-hand side of the Einstein's field equations as in EMSG and hence would provide this interesting feature naturally.}. For instance, as we shall do in what follows, we can set the coupling constants to null for the conventional sources such as the baryons and photons so that these couple to spacetime exactly as in the GR, while we can set a nonzero coupling constant for the sources such as the CDM and relativistic relics (neutrinos) so that these sources couple to the spacetime in accordance with the scale-independent EMSG theory. (iii) Even if we assume that the CDM with $p=0$ and relativistic relics (neutrinos) with $p/\rho=1/3$ are minimally interacting, the violation of the local conservation of the EMTs, as a feature of this theory, the continuity equations come with extra terms like nonminimal interaction terms (which may be dubbed as \textit{pseudo nonminimal interaction}) that lead to new redshift dependencies of the energy densities of these sources to deviate from $a^{-3}$ for CDM and $a^{-4}$ for relativistic relics. Modified redshift dependencies of these sources, which are observationally well motivated and investigated, have been mostly obtained from the non-minimal interaction between dark matter and relativistic relics (see Refs. \cite{DM_nu03,DM_nu04} and references therein). (iv) Another direct consequence of this scenario is that the dimensionless free parameter of the theory $\alpha$ may induce changes on $N_{\rm eff}$ \footnote{$N_{\rm eff}$ quantifies the effective number of species, which via theoretical calculations is well determined within the framework of the standard model, $N_{\rm eff} = 3.046$. The evidence of any positive deviation from this value can be signaling that the radiation content of the Universe is not only due to photons and neutrinos, but also to some extra relativistic relics, the so-called dark radiation and parametrized by $\Delta N_{\rm eff} = N_{\rm eff} - 3.046$. See Ref. \cite{Neff01} for recent observational constraints and Ref. \cite{Neff02} for review.}, which are not due to some extra relativistic species but only as a direct consequence of the pseudo nonminimal interactions of CDM and relativistic relics present in the standard model. 

In order to quantify the observational consequences of the cosmological scenario developed in this study, we use the cosmic microwave background (CMB) Planck data (temperature, polarization, and lensing power spectrum) and baryonic acoustic oscillations (BAO) data. We find that the induced corrections on CDM and relativistic species via the pseudo nonminimal interactions are minimal with $\alpha\sim10^{-7}$, and consequently the deviations from the minimal $\Lambda$CDM model are not observed with statistical significance. We find that the model presented here is quite rich with promising cosmological applications/implications that deserve further investigation.


\section{Scale-independent energy-momentum squared gravity}
 \label{sec:sEMPGmodel}
We consider a modification of the form $f\left(T_{\mu\nu}T^{\mu\nu}, \mathcal{L}_{\rm m}\right)$ in the EH action with a cosmological constant $\Lambda$ which can be written as
\begin{equation}
\begin{aligned}
S=&\int {\rm d}^4x \sqrt{-g}\,\left[\frac{1}{2\kappa}\left(R-2\Lambda \right)+f\left(T_{\mu\nu}T^{\mu\nu}, \mathcal{L}_{\rm m}\right)\right],
\label{action}
\end{aligned}
\end{equation}
where $R$ is scalar curvature, $g$ is the determinant of the metric, and $\mathcal{L}_{\rm m}$ is the Lagrangian density corresponding to the matter source that will be described by the EMT, $T_{\mu\nu}$. Here, the cosmological constant $\Lambda$ is considered as a bare cosmological constant in accordance with the Lovelock's  theorem\footnote{Lovelock's theorem \cite{Lovelock:1971yv,Lovelock:1972vz} states that the only possible second-order Euler-Lagrange expression obtainable in a four-dimensional space from a scalar density of the form $\mathcal{L}= \mathcal{L}(g_{\mu\nu})$ is $E_{\mu\nu}=\sqrt{-g}\left(\lambda_1 G_{\mu\nu}+\lambda_2 g_{\mu\nu}\right)$, where $\lambda_1$ and $\lambda_2$ are constants, leading to Newton's gravitational constant $G$ and cosmological constant $\Lambda$ in Einstein's field equations $G_{\mu\nu}+\Lambda g_{\mu\nu}=\kappa T_{\mu\nu}$ (see Refs. \cite{Bull:2015stt,Clifton:2011jh,Straumann} for further reading).}, stating that $\Lambda$ arises as a constant of nature like Newton's gravitational constant $G=\frac{\kappa}{8\pi}$. 

We vary the action \eqref{action} with respect to the inverse metric as
\begin{equation}
\begin{aligned}
\delta S=\int\, {\rm d}^4 x \sqrt{-g} \bigg[&\frac{1}{2\kappa}\delta R+\frac{\partial f}{\partial(T_{\mu\nu}T^{\mu\nu})}\frac{\delta(T_{\sigma\epsilon}T^{\sigma\epsilon})}{\delta g^{\mu\nu}}\delta g^{\mu\nu}\\
&+\frac{\partial f}{\partial \mathcal{L}_{\rm m}}\frac{\delta \mathcal{L}_{\rm m}}{\delta g^{\mu\nu}}\delta g^{\mu\nu}-\frac{1}{2}g_{\mu\nu}\delta g^{\mu\nu} \\
&\times \bigg\{\frac{1}{2\kappa}\left(R-2\Lambda\right)+f\left(T_{\sigma\epsilon}T^{\sigma\epsilon}, \mathcal{L}_{\rm m}\right)\bigg\}\bigg],
  \end{aligned} 
     \end{equation}
  and, as usual, define the EMT as
  \begin{align}
  \label{tmunudef}
 T_{\mu\nu}=-\frac{2}{\sqrt{-g}}\frac{\delta(\sqrt{-g}\mathcal{L}_{\rm m})}{\delta g^{\mu\nu}}=g_{\mu\nu}\mathcal{L}_{\rm m}-2\frac{\partial \mathcal{L}_{\rm m}}{\partial g^{\mu\nu}},
 \end{align}
which depends only on the metric tensor components and not on its derivatives. We proceed with a particular form of the model,
\begin{align}
\label{function}
f(T_{\mu\nu}T^{\mu\nu},\mathcal{L}_{\rm m})=\sum_i  \left(\alpha_i\sqrt{T_{\mu\nu}^{(i)}T^{\mu\nu}_{(i)}}+\mathcal{L}_{\rm m}^{(i)}\right),
\end{align} 
where $i$ stands for the $i{\rm th}$ fluid. Note that the summation over index $i$ avoids the cross-terms involving product of energy densities when there are at least two sources. In what follows, this ensures the scale independence of the EMSG in the sense that the energy density terms arising due to EMSG combine with the energy density terms arising from GR without changing their scales or powers. The action now reads
\begin{equation}
\begin{aligned}
S=\int {\rm d}^4x \sqrt{-g}\,\bigg[&\frac{R-2\Lambda}{2\kappa}+\sum_i \left(\alpha_i \sqrt{T_{\mu\nu}^{(i)}T^{\mu\nu}_{(i)}}+\mathcal{L}_{\rm m}^{(i)}\right)\bigg],
\label{eq:action}
\end{aligned}
\end{equation}
where $\alpha_i$ ($i=1,2,..,n$) are constants that would take part in  determining the coupling strength of the EMSG modifications to gravity and $n$ is the number of different fluids under consideration. We note that the monofluid case of this model for general $\eta$ has been recently studied in Refs. \cite{Akarsu:2017ohj,Board:2017ign}. 
In this work, we study the gravity model under consideration in the context of multifluid cosmology. From the action given in \eqref{eq:action}, we reach the following modified Einstein's field equations,
  \begin{equation}
  \begin{aligned}
&G_{\mu\nu}+\Lambda g_{\mu\nu} \\
&=\kappa \sum_i \Bigg[T_{\mu\nu(i)}+\alpha_i \sqrt{T_{\sigma\epsilon(i)}T^{\sigma\epsilon(i)}} \left(g_{\mu\nu}-\frac{\Xi_{\mu\nu(i)}}{T_{\sigma\epsilon(i)}T^{\sigma\epsilon}_{(i)}}\right)\Bigg],
\label{fieldeq}
\end{aligned}
\end{equation} 
where $G_{\mu\nu}=R_{\mu\nu}-\frac{1}{2}Rg_{\mu\nu}$ is the Einstein tensor and the new tensor is defined as\footnote{The EMT given in \eqref{tmunudef} does not include the second variation of $\mathcal{L}_{\rm m}$; hence, the last term of \eqref{theta0} is null. As it is well known that the definition of matter Lagrangian giving the perfect fluid EMT is not unique, one could choose either $\mathcal{L}_{\rm m}=p$ or $\mathcal{L}_{\rm m}=-\rho$, which provides the same EMT (see Refs. \cite{Bertolami:2008ab,Faraoni:2009rk} for a detailed discussion). In the present study, we consider $\mathcal{L}_{\rm m}=p$. }
\begin{equation}
\begin{aligned}
\Xi_{\mu\nu(i)}=&-2\mathcal{L}_{\rm m(i)}\left(T_{\mu\nu(i)}-\frac{1}{2}g_{\mu\nu}T_{(i)}\right)-T_{(i)}T_{\mu\nu(i)} \\
&+2T_{\mu\,\,(i)}^{\gamma}T_{\nu\gamma(i)}-4T^{\sigma\epsilon(i)}\frac{\partial^2 \mathcal{L}_{{\rm m}(i)}}{\partial g^{\mu\nu} \partial g^{\sigma\epsilon}}.
\label{theta0}
\end{aligned}
\end{equation}

Taking the covariant derivative of \eqref{fieldeq}  and raising an index with the metric, we obtain the local conservation of EMT as follows:
\begin{equation}
\begin{aligned}
\label{covdertmunu}
\sum_i\Bigg\{\bigg[&\nabla_{\mu}T^{\mu}_{\,\,\nu(i)}+\alpha_i \bigg\{\left( \frac{\delta^{\mu}_{\,\,\nu}}{2}+\frac{\Xi^{\mu}_{\,\,\nu(i)}}{2 T^{\sigma}_{\epsilon(i)}T^{\epsilon}_{\,\,\sigma(i)}}\right) \\
&\times \frac{\partial_{\mu} \left(T^{\sigma}_{\,\,\epsilon(i)}T^{\epsilon}_{\,\,\sigma(i)}\right)}{\sqrt{T^{\sigma}_{\,\,\epsilon(i)}T^{\epsilon}_{\,\,\sigma(i)}}}-\frac{\nabla_{\mu}\Xi^{\mu}_{\,\,\nu(i)}}{\sqrt{T^{\sigma}_{\,\,\epsilon(i)}T^{\epsilon}_{\,\,\sigma(i)}}}\bigg\}\bigg]\Bigg\}=0.
\end{aligned}
\end{equation}
Note that the local/covariant energy-momentum conservation $\nabla^{\mu}T_{\mu\nu(i)}=0$ is not satisfied for $\alpha_i\neq0$ in general.

\section{Cosmology}
\label{sec:backg}
We consider the spatially maximally symmetric spacetime metric, i.e., the Robertson-Walker (RW) metric, with flat spacelike sections given as
\begin{align}
\label{RW}
{\rm d}s^2=-{\rm d}t^2+a^2\,({\rm d}x^2+{\rm d}y^2+{\rm d}z^2),
\end{align}  
where the scale factor $a=a(t)$ is a function of cosmic time $t$ only, and the perfect fluid forms of the EMTs are given by
\begin{align}
\label{em}
T_{\mu\nu(i)}=(\rho_{i}+p_i)u_{\mu}u_{\nu}+p_i g_{\mu\nu},
\end{align} 
where $\rho_i$ and $p_i$ are, respectively, the energy density and the pressure of the $i{\rm th}$ fluid and $u_{\mu}$ is the four-velocity of the medium satisfying the conditions $u_{\mu}u^{\mu}=-1$ and $\nabla_{\nu}u^{\mu}u_{\mu}=0$. 

Using \eqref{em} and a barotropic equation of state (EoS) $w_i=\frac{p_i}{\rho_i}={\rm const.}$, we find $\Xi_{\mu\nu(i)}$ given 
in \eqref{theta0} and the self-contraction of the EMT as 
\begin{align}
\label{thetafrw}
\Xi_{\mu\nu(i)}=&-\rho_i^2(3w_i+1)(w_i+1)u_{\mu}u_{\nu},\\
T_{\mu\nu(i)}T^{\mu\nu(i)}=&\,\rho_i^2(3w_i^2+1),
\label{TTfrw}
\end{align}
respectively. The local energy conservation equation \eqref{covdertmunu} reads
\begin{equation}
\begin{aligned}
\label{noncons}
 \sum_i \Bigg[&\bigg\{\dot \rho_i+ 3H(1+w_i)\rho_i\bigg\}+\alpha_i\Bigg\{\frac{4w_i}{\sqrt{3w_i^2+1}}\dot \rho_i\\
 &+3H\rho_i\left(\frac{3w_i^2+4w_i+1}{\sqrt{3w_i^2+1}}\right)\Bigg\}\Bigg]=0,
\end{aligned}
\end{equation}
where $H=\frac{\dot{a}}{a}$ is the Hubble parameter. We consider that the Universe is filled with four different types of sources: CDM, baryons, photons, and neutrinos plus any possible relativistic relics, with energy densities $\rho_{\rm cdm}$, $\rho_{\rm b}$, $\rho_{\gamma}$, and $\rho_{\nu}$, respectively. We assume baryons and CDM with EoS $w_{\rm b}=w_{\rm cdm}=0$ and photons and neutrinos plus any possible relativistic relics with EoS $w_{\gamma}=w_{\nu}=\frac{1}{3}$ and accordingly calculate the relevant tensors given in \eqref{thetafrw} and \eqref{TTfrw}. Note that we break the EP for CDM particles and neutrinos plus any possible relativistic relics such that we assume that standard fluids, i.e., photons ($\gamma$) and baryons (b), couple to gravity in the same way as in the standard GR; i.e., for these fluids, the corresponding EMSG modification is considered to be null by choosing $\alpha_{\gamma}=\alpha_{\rm b}=0$, while the neutrinos plus any possible relativistic relics and CDM particles couple to the spacetime in accordance with the EMSG modification with $\alpha_{\nu}\neq0\neq\alpha_{\rm cdm}$. In view of Occam's razor, we choose $\alpha_{\nu}=\alpha_{\rm cdm}=\alpha$ to reduce the number of free parameters in the model \footnote{If one allows $\alpha_{\nu}$ and $\alpha_{\rm cdm}$ to be two separate coupling parameters, we expect a statistical degeneracy among them, as the current data are not capable of breaking the degeneracy among the two different phenomenological parameters with the same physical nature (in our case, the coupling parameters $\alpha_{\nu}$ and $\alpha_{\rm cdm}$). Thus, it is reasonable to maintain the same coupling parameter for the CDM particles and neutrinos plus any possible relativistic relics.}. Besides, note that, even for the choice $\alpha_{\nu}=\alpha_{\rm cdm}=\alpha$, the gravitational coupling in the dark sector is species dependent because of the different values of EoS parameters of CDM and relativistic relics.

Regarding the violation of the EP in our model, in fact it is not uncommon in fundamental physics; for instance, it is intimately connected with some of the basic aspects of the unification of gravity with particle physics such as string theories (see Ref. \cite{Uzan:2010pm} and references therein). Moreover, for the dark sector, it has been recently studied that the validity of EP is questionable and may be violated \cite{peebles10}.

Note that we do not count $\Lambda$ among the sources since in this study we consider $\Lambda$ as a bare cosmological constant rather than the conventional vacuum energy described by the EoS parameter equal to minus unity; i.e., we assume that conventional vacuum energy is null \footnote{If we consider conventional vacuum energy with $p_{\rm vac}=-\rho_{\rm vac}$, it would contribute to the right-hand side of \eqref{eq:rhoprime} and \eqref{eq:presprime} as $\kappa(1-2\alpha)\rho_{\rm vac}$ and $-\kappa(1-2\alpha)\rho_{\rm vac}$, respectively.}. Although we know of no special symmetry that could enforce a vanishing vacuum energy while remaining consistent with the known laws of physics. It is usually thought to be easier to imagine an unknown mechanism that would set vacuum energy precisely to zero than one that would suppress it by just the right amount $\rho_{\rm vac}^{\rm (observation)}/\rho_{\rm vac}^{(\rm theory)}\sim10^{-120}$ to yield an observationally accessible vacuum energy (see Refs. \cite{Weinberg:1988cp,Padmanabhan:2002ji} and references therein).

Using the relevant tensors in the field equations \eqref{fieldeq} in the RW metric given in \eqref{RW}, we obtain the expansion rate of the Universe given by the Hubble parameter,
\begin{align}
3H^2 =  \Lambda + \kappa \Big[ \rho_{\rm \gamma} + \rho_{\rm b} + \rho_{\rm cdm} + \left(1+\frac{2\alpha}{\sqrt{3}}\right) \rho_{\nu} \Big], \label{eq:rhoprime}
\end{align}
and the pressure equation,
\begin{align}
-2\dot{H}-3H^2=-\Lambda+\kappa\left[\frac{1}{3}\rho_{\gamma}+\alpha\rho_{\rm cdm}+\left(\frac{1}{3}+\frac{2\alpha}{\sqrt{3}}\right)\rho_{\nu}\right].
\label{eq:presprime}
\end{align}

We find from \eqref{noncons} that the joint continuity equation for the CDM and relativistic relics is
\begin{equation}
\begin{aligned}
\label{noncons1}
&\dot \rho_{\rm cdm}+ 3H\rho_{\rm cdm}+\dot \rho_{\nu}+ 4H\rho_{\nu} \\
&+\alpha\Bigg( 3H \rho_{\rm cdm}+\frac{2}{\sqrt{3}}\dot{\rho}_{\nu}+4\sqrt{3}H\rho_{\nu} \Bigg)=0,
\end{aligned}
\end{equation}
while it is as usual for the standard fluids giving rise to $\rho_{\rm b}\propto a^{-3}$ and $\rho_{\rm \gamma}\propto a^{-4}$ for baryons and photons, respectively. To close the system, it is reasonable to assume that these two fluids are minimally interacting (i.e., interacting only gravitationally), which would lead to the separation of \eqref{noncons1} into two pieces as follows:
\begin{equation}
\begin{aligned}
\label{rhomint1}
\dot \rho_{\rm cdm}+ 3H\rho_{\rm cdm}=&-3\alpha H \rho_{\rm cdm}, 
\end{aligned}
\end{equation}
\begin{equation}
\begin{aligned}
\label{rhomint2}
\dot \rho_{\nu}+ 4H\rho_{\nu}=&-\frac{4\alpha}{2\alpha+\sqrt{3}} H \rho_{\nu}.
\end{aligned}
\end{equation}
It is noteworthy that although we assume that these two sources are minimally interacting, each continuity equation comes with an extra term on the right-hand side that appears like a nonminimal interaction term, which may be interpreted as ``pseudo nonminimal interaction" of each source. These terms appear from nonconservation of EMTs, and it is not necessary for them to cancel each other in contrast to a general relativistic model considered with two nonminimally interacting sources. Finally, solving the continuity equations \eqref{rhomint1} and \eqref{rhomint2}, we reach the following background evolution equations for CDM and relativistic relics sources,
\begin{equation}
\rho_{\rm cdm}= \rho_{{\rm cdm},0}\,a^{-3-3\alpha}=\rho_{{\rm cdm},0}\,(1+z)^{3+3\alpha},
\end{equation}
\begin{equation}\label{rhomint3}
\rho_{\nu}=\rho_{{\nu},0}\,a^{-4-\frac{4\alpha}{2\alpha+\sqrt{3}}}=\rho_{{\nu},0}\,(1+z)^{4+\frac{4\alpha}{2\alpha+\sqrt{3}}},
\end{equation}
where $z=-1+1/a$ is the redshift. We note that EMSG corrections controlled by the constant $\alpha$ modify the redshift dependence of energy densities like in the nonminimally interacting models in GR; namely, the power of the redshift dependencies are modified by $\alpha$. 

\section{Scalar perturbations}
\subsection{General equations}
In this section, we derive the general form of the equations which describe small cosmological perturbations within the context of the scale-independent EMSG theory. In conformal Newtonian (longitudinal) gauge, the line element has the form
\begin{equation}
\begin{aligned}
\label{pertrw}
{\rm d}s^2= a(\eta)^2\left[-\left(1+2\psi\right){\rm d}\eta^2+\left(1-2\phi\right) \delta_{ij} {\rm d}x^{i}{\rm d}x^{j}\right],
\end{aligned}
\end{equation}
where $\eta$ is the conformal time and $\psi(\eta)$ and $\phi(\eta)$ are metric potentials, in the longitudinal gauge, which coincide with the Bardeen potentials. A perturbation in the metric must be matched with a perturbation in the EMT (the matter or energy that occupies that space), and hence perturbations on energy densities and pressures are given as
\begin{equation}
\label{rhop_pert}
\rho_i = \rho_i+\delta\rho_i \quad \textnormal{and}\quad p_i =p_i+\delta p_i,
\end{equation}
where $\delta_i=\frac{\delta \rho_i}{\rho_i}$ and the speed of sound is $c_{s,i}^2=\frac{\delta p_i}{\delta \rho_i}$. The total four-velocity can be written as a perturbation to this nonperturbed velocity $u^{\mu}=a^{-1}\delta^{\mu}_{0}+\delta u^{\mu}$, which must satisfy the four-velocity identity, $u_{\mu}u^{\mu}=-1$, which puts a constraint on the perturbed components. The total four-velocity reads $u_0=a^{-1}(1-\psi)$ and $u_i=\delta u_i$.

Substituting \eqref{rhop_pert} and total four-velocity, given just above, into \eqref{em} with the barotropic EoS $w_i=\frac{p_i}{\rho_i}$, and collecting the terms up to the first-order perturbations, we obtain
\begin{equation}
\begin{aligned}
 T_i=&-\rho_i\left[(1-3w_i)+\delta_f(1-3c_{s,i}^2)\right], \\
 \left(T^{\sigma}_{\epsilon}T^{\epsilon}_{\,\,\sigma}\right)_i  =&\,\rho_i^2\left[(1+3w_i^2)+2\delta_i(1+3w_i c_{s,i}^2)\right],
  \end{aligned}
\end{equation}
respectively.
In a similar way, the new tensor $\Xi$ given in \eqref{thetafrw} becomes 
\begin{equation}
\begin{aligned}
\Xi^{\mu}_{\nu(i)}=-\rho_i^2\Bigg[&(3w_i^2+4w_i+1) \left\{u^{\mu}\left(u_{\nu}+\delta u_{\nu}\right)+u_{\nu} \delta u^{\mu}\right\}\\
&+2\delta_i\left\{3w_i c_{s,i}^2+2(w_i+c_{s,i}^2)+1\right\} u^{\mu}u_{\nu} \Bigg].
 \end{aligned}
\end{equation}
Substituting all the perturbed quantities (up to the first order) calculated above into \eqref{covdertmunu} and considering $\nu=0$, we obtain the background continuity equation given in \eqref{noncons} from the zeroth-order terms, while the first-order terms yield the perturbed part of the continuity equation as
\begin{equation}
\begin{aligned}
\label{contpert}
 \sum_i \Bigg[ &\delta'_{i} \bigg(1+4\alpha_i \frac{c_{s,i}^2+3w_i^3}{(1 + 3 w_i^2)^{\frac{3}{2}}}\bigg)\\ 
 &+(\theta_i-3 \phi')\left(1+w_i+\alpha_i \frac{3w_i^2+4w_i+1}{\sqrt{1+3w_i^2}}\right)  \\
&+3 \mathcal{H}\delta_i \bigg\{1+c_{s,i}^2+\alpha_i \bigg(\frac{3w_i c_{s,i}^2+1}{\sqrt{1+3w_i^2}}+4\frac{c_{s,i}^2+3w_i^3}{(1+3w_i^2)^{\frac{3}{2}}}\bigg) \\
&-\frac{1+4\alpha_i \frac{c_{s,i}^2+3w_i^3}{(1 + 3 w_i^2)^{\frac{3}{2}}}}{1+4\alpha_i \frac{w_i}{\sqrt{1 + 3 w_i^2}}}\bigg(1+w_i+\alpha_i \frac{3w_i^2+4w_i+1}{\sqrt{1+3w_i^2}}\bigg)\bigg\} \Bigg]\\
& = 0.
\end{aligned}
\end{equation}
Now we consider the $\nu=i$\,-component of \eqref{covdertmunu}; there is no background equation when the zeroth-order terms are taken, since $T^{\mu}_{i}=0$. We have $u^{i}=0$; therefore, $\Xi^{i}_{j}=\Xi^{i}_{i}=0$, $T^{i}_{0}=0$ and $\delta T^{0}_{i}=\theta$, where $\theta$ is the divergence of velocity. The perturbed Euler equation reads
\begin{align}
\label{eulerpert}
\sum_i \Bigg[ &\left\{ \theta_i'+\mathcal{H}\theta_i \left(1 - 3 w_{i}+3\alpha_i\frac{w_i^2-1}{\sqrt{1+3w_i^2}+4\alpha_i w_i}\right)\right\}  \nonumber \\
&\times \left(1-\alpha_i\frac{3w_i^2+4w_i+1}{(1+w_i)\sqrt{1 + 3 w_i^2}}\right) \nonumber \\ 
&- \frac{c_{s,i}^2-\alpha_i\sqrt{1 + 3 w_i^2} }{1+w_i} k^2 \delta_{i}  \Bigg]- k^2 \psi =0.
\end{align}
\subsection{CDM and relativistic relics perturbations}
We now write the perturbation equations derived above for the cosmological model with $\alpha_{\gamma}=\alpha_{\rm b}=0$ and $\alpha_{\nu}=\alpha_{\rm cdm}=\alpha$, described  in Sec.  \ref{sec:backg}. For relativistic relics ($i= {\nu}$), we have $w_{\nu}=c_{s,\nu}^2=1/3$. Therefore, Eqs. \eqref{contpert} and \eqref{eulerpert} reduce to
\begin{align}
\label{contdr}
\delta'_{\nu}+\frac{4+4\sqrt{3}\alpha}{3+2\sqrt{3}\alpha}\left(\theta_{\nu}-3 \phi'\right)= 0,
\end{align}
\begin{equation}
\begin{aligned}
\label{eulerdr}
\left(1-\sqrt{3}\alpha\right)\left(\theta'_{\nu}-
\frac{4\alpha}{2\alpha+\sqrt{3}}\mathcal{H}\theta_{\nu}\right)&\\
-\left(\frac{1-2\sqrt{3}\alpha}{4}\right)k^2 \delta_{\nu} 
- k^2 \psi&=0,
\end{aligned}
\end{equation}
respectively. For CDM ($i={\rm cdm}$), we have $w_{\rm cdm}=c_{s,\rm cdm}^2=0$. Therefore, the first-order continuity and Euler equations for CDM from (\ref{contpert}) and \eqref{eulerpert} read as
\begin{align}
\label{contdm}
\delta'_{\rm cdm}+(1+\alpha)\left(\theta_{\rm cdm}-3 \phi'\right)= 0,
\end{align}
\begin{align}
\label{eulerdm}
(1-\alpha)\left[\theta'_{\rm cdm}+\left(1-3\alpha\right)\mathcal{H}\theta_{\rm cdm}\right]+\alpha k^2 \delta_{\rm cdm}- k^2 \psi=0,
\end{align}
respectively. The Boltzmann hierarchy for the relativistic relics follows the standard procedures as described in Ref. \cite{Ma_Bertschinger} (see also Ref. \cite{class}).

\section{Observational constraints}
\label{obs.contraints}
\subsection{Model parameters, data sets and methodology}

In what follows, we consider that the relativistic relics are the three species of neutrinos (two massless and one massive) predicted by the standard model, as assumed by the Planck team \cite{Planck2015}. Also, we assume a normal mass hierarchy and fix the mass scale to 0.06 eV. The effective number of species can be parametrized, when the neutrinos are relativistic, by
\begin{eqnarray}
 N_{\rm eff} =  \frac{8}{7} \left(\frac{4}{11} \right)^{-\frac{4}{3}} \left(\frac{\rho_{\nu}}{\rho_{\gamma}} \right).
\end{eqnarray}

From (\ref{rhomint3}), we can see that $\alpha$ corrections on $\rho_{\nu}$ will naturally induce changes on $N_{\rm eff}$, depending on the redshift as well. As we expect corrections like $\alpha \lll 1$ in this study, possible new dependencies at high redshifts are minimal. Also due to instability at early times, we eliminate the dependencies of $N_{\rm eff}$ in $z$ in our implementation and evaluate the effective number of (free-streaming) relativistic species at present time. Note that changes in $N_{\rm eff}$ can also take place due to some other phenomenon like leptonic asymmetry (see Refs.\cite{xi01, xi02} and \cite{Neff01} for recent observational results), the low-reheating scenario \cite{Salas, Kawasaki}, and nonstandard interactions between neutrinos and electrons \cite{Mangano}. Thus, the cosmological scenario proposed here can also induce changes on $N_{\rm eff}$ that are not due to some new extra species (dark radiation) but due to the nonstandard dynamics of the relativistic species induced by the scale-independent EMSG. So, we choose $N_{\rm eff}$ as a free parameter in our analyses. Therefore, in addition to the baseline present in the $\Lambda$CDM model, the cosmological scenario proposed here has $\alpha$ and $N_{\rm eff}$ as additional free parameters. Finally, the base parameter set for our  model is
 \[
 P = \{\omega_{\rm b}, \, \omega_{\rm cdm}, \, 100\theta_{s}, \, \ln10^{10}A_{s}, \,  
 n_s, \, \tau_{\rm reio}, \, \alpha,  \, N_{\rm{eff}} \},
 \]
where the first six parameters are the base parameters of the minimal $\Lambda$CDM model \cite{p13}.

In order to constrain the free parameters of the model, we use the CMB data from Planck 2015 with the likelihoods at multipoles $l \geq 30$ using TT, TE, and EE power spectra and the low-multipole polarization likelihood at $l \leq 29$ in addition to the Planck 2015 CMB lensing power spectrum likelihood \cite{Planck2015}. We also use the BAO measurements from   (i) the Six  Degree  Field  Galaxy  Survey \cite{bao1}, 
(ii) the  Main  Galaxy  Sample  of  Data  Release 7  of  Sloan  Digital  Sky  Survey \cite{bao2}, 
(iii) the  LOWZ  and  CMASS  galaxy  samples  of  the Baryon  Oscillation  Spectroscopic  Survey (BOSS) \cite{bao3},  and (iv) the distribution of the LymanForest in the BOSS \cite{bao4}. These BAO data points are summarized in Table I of Ref. \cite{coupled04}.

We have modified the publicly available CLASS \cite{class} code for the model under consideration, and interfaced it with the MONTE PHYTON \cite{monte} code using the Metropolis-Hastings algorithm with uniform priors on the model parameters. Then, the correlated Markov Chain Monte Carlo samples are obtained by considering two combinations of data sets: CMB and CMB + BAO, where the convergence of the Monte Carlo Markov Chains for all the model parameters is ensured according to the Gelman-Rubin criteria \cite{Gelman}. The output samples are analyzed by using the GETDIST PYTHON package \cite{antonygetdist}.

\subsection{Results and discussions}
\label{sec:obsresults}

\begin{figure}[!ht]\centering
\captionsetup{justification=raggedright,singlelinecheck=false}
\includegraphics[width=0.43\textwidth]{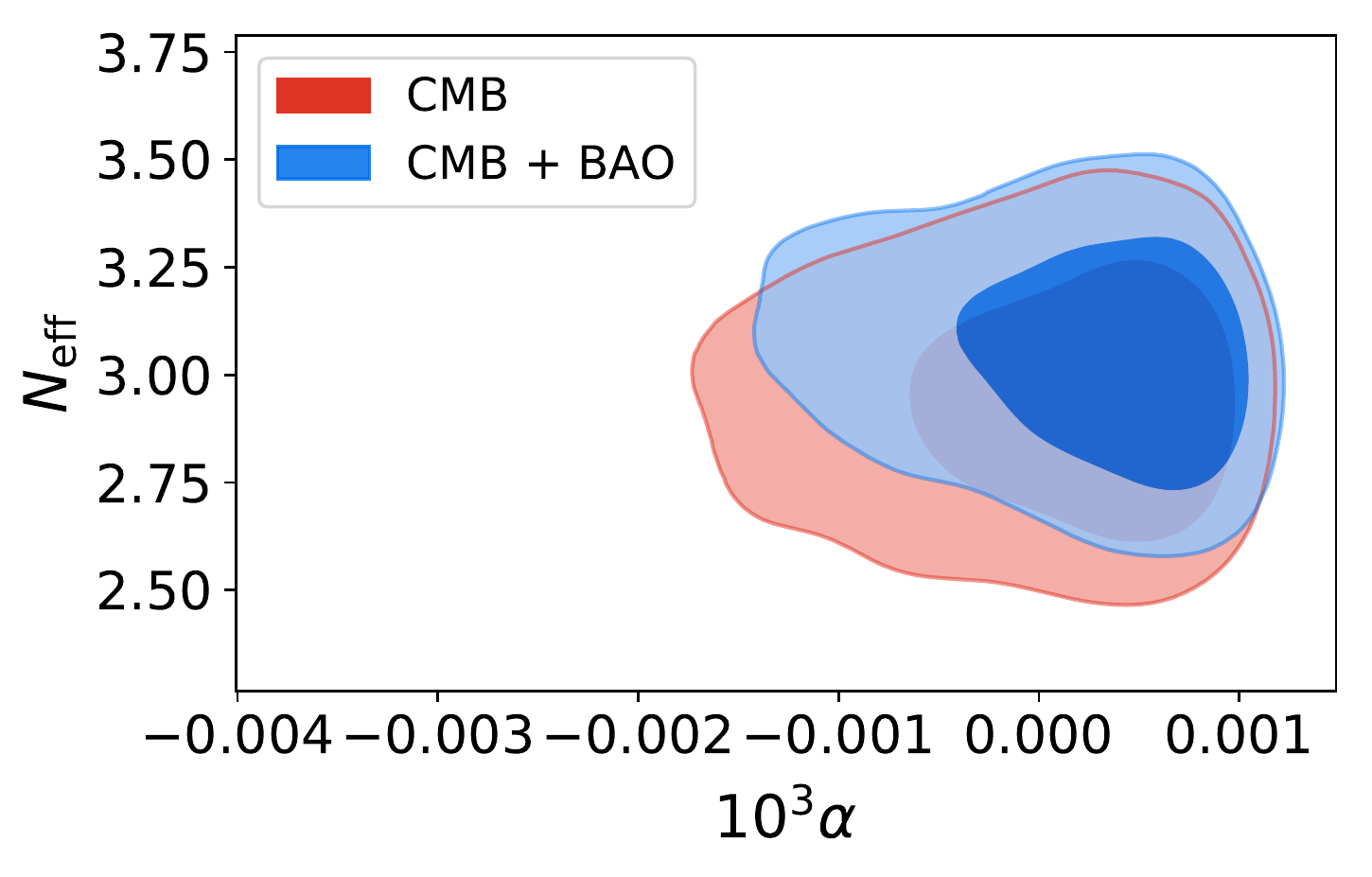}
\caption{\label{fig1} Parametric space in the plane $\alpha - N_{\rm eff}$ from CMB and CMB + BAO data.}
\end{figure}

Table \ref{tab:results} summarizes the observational constraints on the free parameters of the model. Figure \ref{fig1} shows the parametric space in the plane $\alpha - N_{\rm eff}$ at 68\% and $95\%$ C.L. We note that negative $\alpha$ values tend to decrease the values of $N_{\rm eff}$. Corrections that decrease the $N_{\rm eff}$ values are also predicted by the low-reheating scenario \cite{Salas, Kawasaki}. It is important to mention that the three species of neutrinos fix $N_{\rm eff} = 3.046$, and deviations from this value may be due to the pseudo nonminimal interactions of CDM and relativistic relics. Obviously, in general, deviation of $N_{\rm eff}$ may be due to both the contributions, that is, some dark radiation (extra relativistic relics) and $\alpha$ corrections (or pseudo nonminimal interactions in CDM and relativistic relics). Denoting these two contributions, respectively, by $\Delta N^{\rm dr}_{\rm eff}$ and $\Delta N^{\alpha}_{\rm eff}$, we can quantify the net deviation as $\Delta N_{\rm eff} = \Delta N^{\alpha}_{\rm eff} + \Delta N^{\rm dr}_{\rm eff}$. There are many models for dark radiation species, which contribute differently to $N_{\rm eff}$, for instance, small contributions like massless dark gluons with $\Delta N^{\rm dr}_{\rm eff} \simeq 0.07$ \cite{Abad}, Goldstone bosons with $\Delta N^{\rm dr}_{\rm eff} \simeq 0.3$ \cite{Weinberg02}, a fully thermalized sterile neutrino with $\Delta N^{\rm dr}_{\rm eff} = 1.0$ \cite{Abazajian}, to name a few.

\begin{table}[!ht]
 \captionsetup{justification=justified,singlelinecheck=false,font=footnotesize}
\begin{center}
\caption{Constraints on the model parameters from CMB and CMB + BAO data. The derived parameter $H_0$ is in the units of km
s${}^{-1}$ Mpc${}^{-1}$.}
\label{tab:results}
\begin{tabular} { l  c c}
\hline
 Parameter &  CMB & CMB + BAO\\
\hline
{$10^2\omega_{b }$} & $2.219\pm 0.024            $&$2.237\pm 0.020            $\\

{$\omega_{\rm{cdm} }$} & $0.1176^{+0.0026}_{-0.0031}$&$0.1179\pm 0.0030          $\\

{$100 \theta_{s }  $} & $1.04210\pm 0.00051        $&$1.04199\pm 0.00050        $\\

{$\ln 10^{10}A_{s }$} & $3.054^{+0.024}_{-0.034}   $&$3.066^{+0.028}_{-0.033}   $\\

{$n_{s }         $} & $0.9613\pm 0.0093          $&$0.9685\pm 0.0075          $\\

{$\tau_{\rm {reio} }   $} & $0.063^{+0.010}_{-0.017}   $&$0.068^{+0.013}_{-0.016}   $\\

{$10^{3}\alpha   $} & $0.00011^{+0.00078}_{-0.00029}$&$0.00028^{+0.00067}_{-0.00024}$\\

{$N_{\rm eff}    $} & $2.94\pm 0.20              $&$3.04\pm 0.18              $\\

{$H_0            $} & $66.9\pm 1.6               $&$68.2\pm 1.3               $\\

\hline
\end{tabular}
\end{center}
\end{table}

Thus, it is reasonable to expect both types of contributions to $\Delta N_{\rm eff}$ in our model. To quantify the decomposition of $\Delta N_{\rm eff}$, it is necessary to specify a prior dark radiation model and then to directly analyze the possible contributions to $\Delta N^{\alpha}_{\rm eff}$. But, without lost of generality, we may also think that any deviation in $N_{\rm eff}$ from its default value can be only due to the pseudo nonminimal interactions of CDM and relativistic relics, and not due to any new species.

\begin{figure}[!ht]\centering
\captionsetup{justification=raggedright,singlelinecheck=false}
\includegraphics[width=0.43\textwidth]{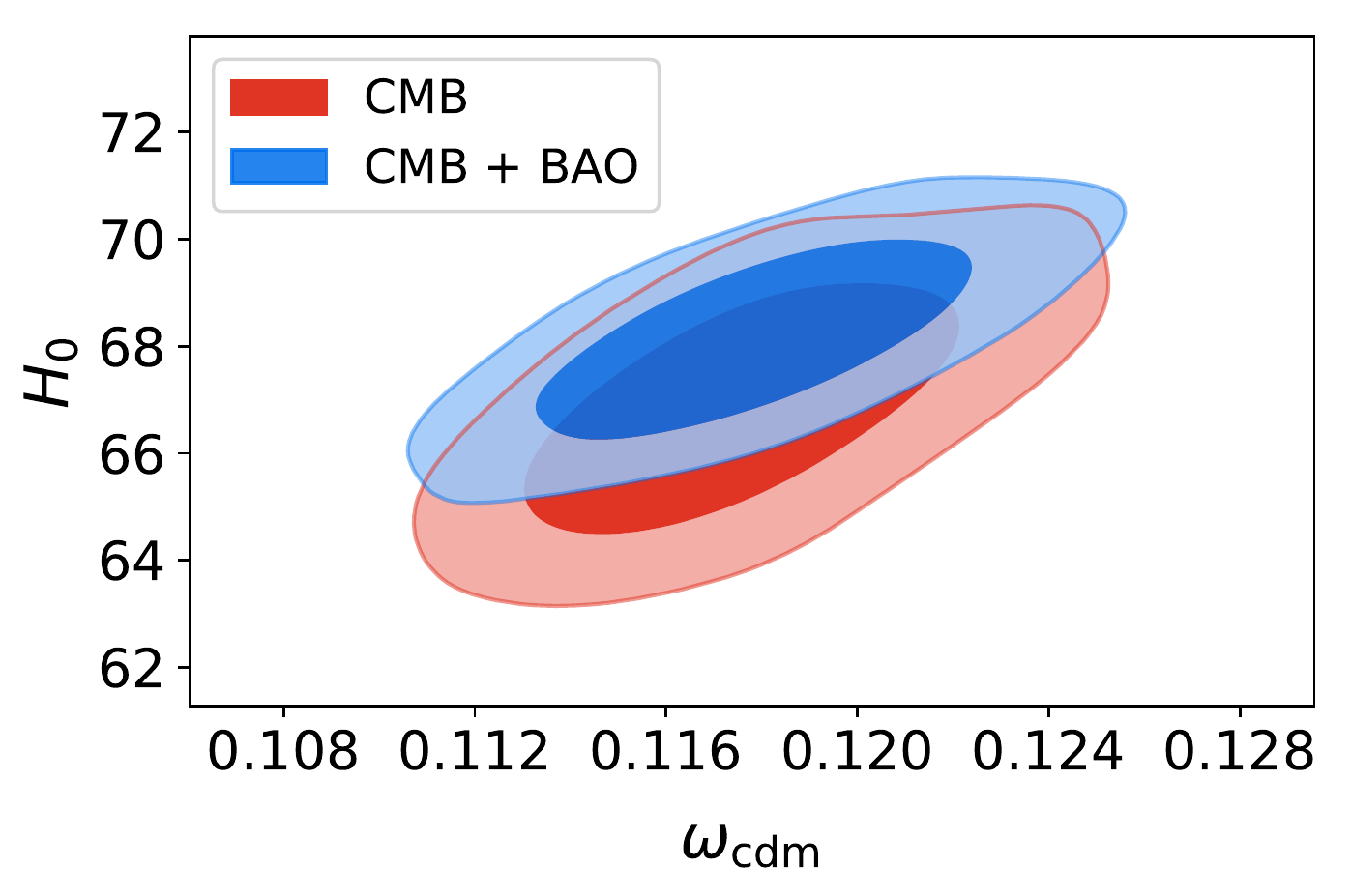}
\caption{\label{fig2} Parametric space in the plane $w_{\rm cdm} - H_0$ from CMB and CMB + BAO data.}
\end{figure}

Figure \ref{fig2} shows the parametric space in the plane $w_{\rm cdm} - H_0$  at 68\% and $95\%$ C.L. The CDM density is modified at both levels (background and perturbative), but we can note that these corrections are minimal, and the constraints on CDM density do not present significant deviations. The case for $H_0$ is similar. Thus, both parameters have similar constraints as those foreseen for the $\Lambda$CDM without corrections. Since $\alpha$ may induce direct changes in $N_{\rm eff}$, it is expected that $\alpha$ could also, in principle, correlate with $H_0$ due to the well-known strong correlation between $N_{\rm eff}$ and $H_0$. Here this is not observed due to the small $\alpha$ values ($\sim 10^{-7}$). But, in general, $\alpha > 0$ (i.e., $\Delta N^{\alpha}_{\rm eff} > 0$) can increase $H_0$ values. The constraints on $\alpha$ are very small, more specifically $\alpha \sim 10^{-7}$, and the constraints obtained from CMB and CMB + BAO on $\alpha$ are very similar to each other. Both analyses are compatible with each other even at 68\% C.L. In fact, the constraints over the entire base parameters of the model are very close to $\Lambda$CDM model, and therefore significant corrections on $\Lambda$CDM model are not observed.

Significant values of $\alpha$ can increase (or decrease) the $N_{\rm eff}$ constraints. If $\alpha$ increases $N_{\rm eff}$, the Universe expands faster during the radiation dominated era, and it causes the Universe to be younger at any given redshift. Also, the comoving sound horizon at recombination will be smaller. But, here, we expect minimal corrections on these quantities. An analysis based on the current tensions in some cosmological data can be carried out in some future communication within the context of the model presented here.

\begin{figure}[h]
\captionsetup{justification=raggedright,singlelinecheck=false}
\par
\begin{center}
\subfigure[]{            \label{fig3a}
            \includegraphics[width=0.41\textwidth]{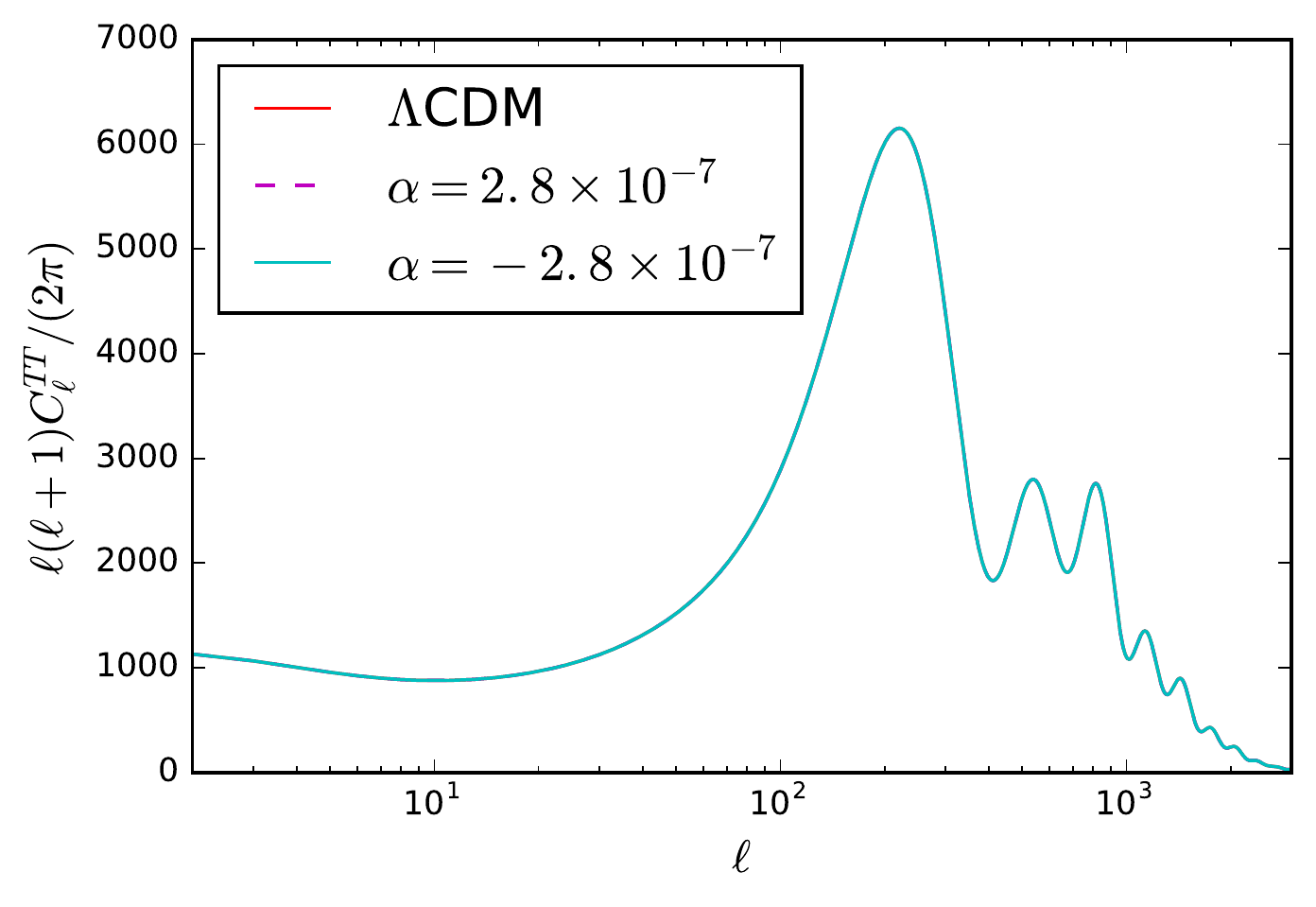}
        }
\subfigure[]{           \label{fig3b}
           \includegraphics[width=0.42\textwidth]{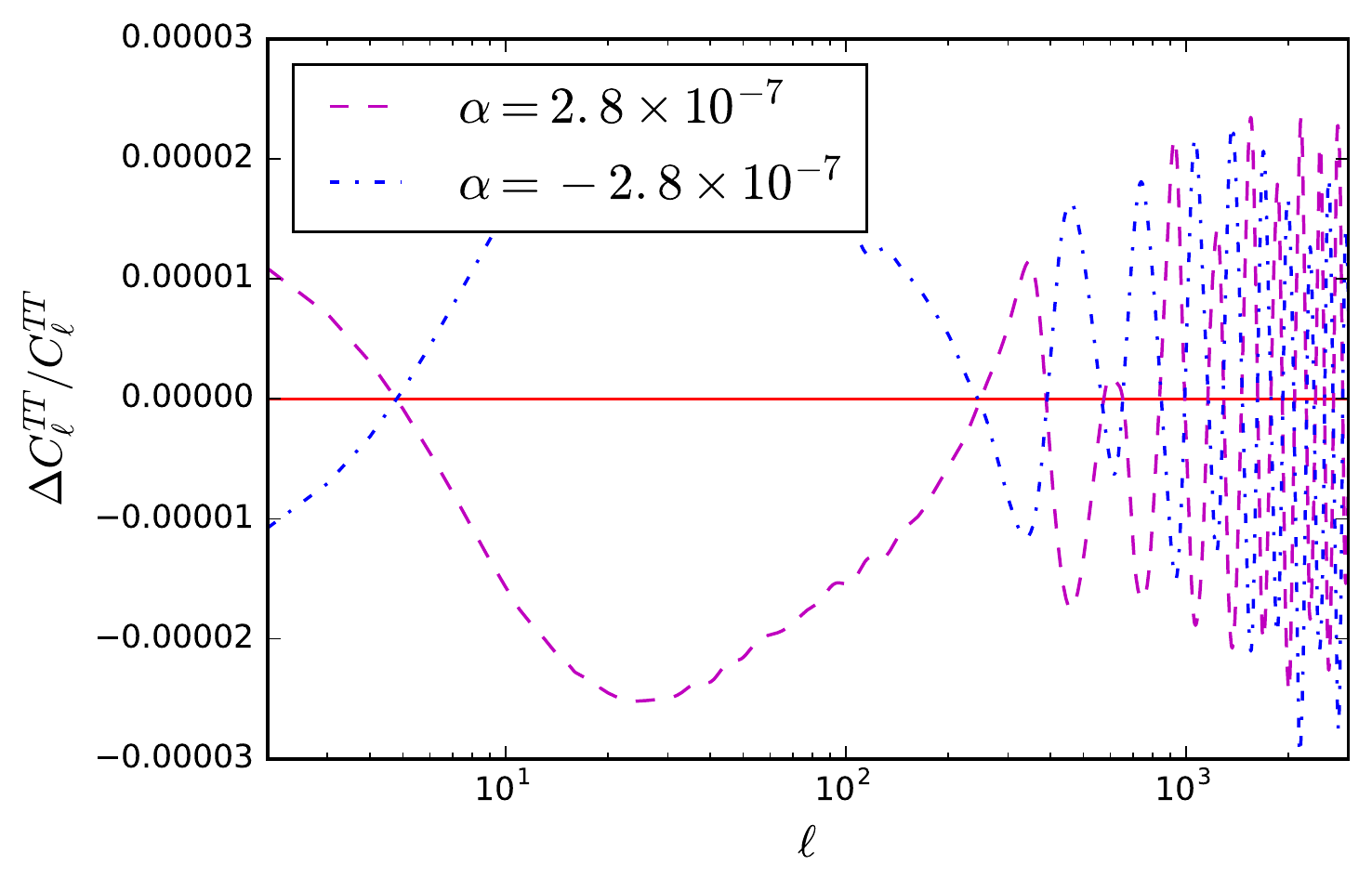}
        }
\end{center}
\caption{ (a) Theoretical prediction for the CMB TT power spectrum in the presence of the pseudo nonminimal interactions in CDM and relativistic relics for different values of $\alpha$, and the $\Lambda$CDM model. (b) Same as in (a), but quantifying deviations with respect to $\Lambda$CDM model. For drawing the graphs, we have chosen the best fit mean values from Table \ref{tab:results} for the baseline
parameters. }
\end{figure}

Figures \ref{fig3a} and  \ref{fig3b} show the theoretical prediction for the CMB TT power spectrum in the presence of the pseudo nonminimal interactions in CDM and relativistic relics for different values of $\alpha$. Due to the very small corrections provided by $\alpha$, deviations from $\Lambda$CDM are minimal, around 0.001 \%, on CMB TT. Originally, other or the larger values of $\alpha$ can be chosen to increase the effects on the CMB power spectrum. 

\begin{figure}[h]
\captionsetup{justification=raggedright,singlelinecheck=false}
\par
\begin{center}
\subfigure[]{             \label{fig4a}
            \includegraphics[width=0.41\textwidth]{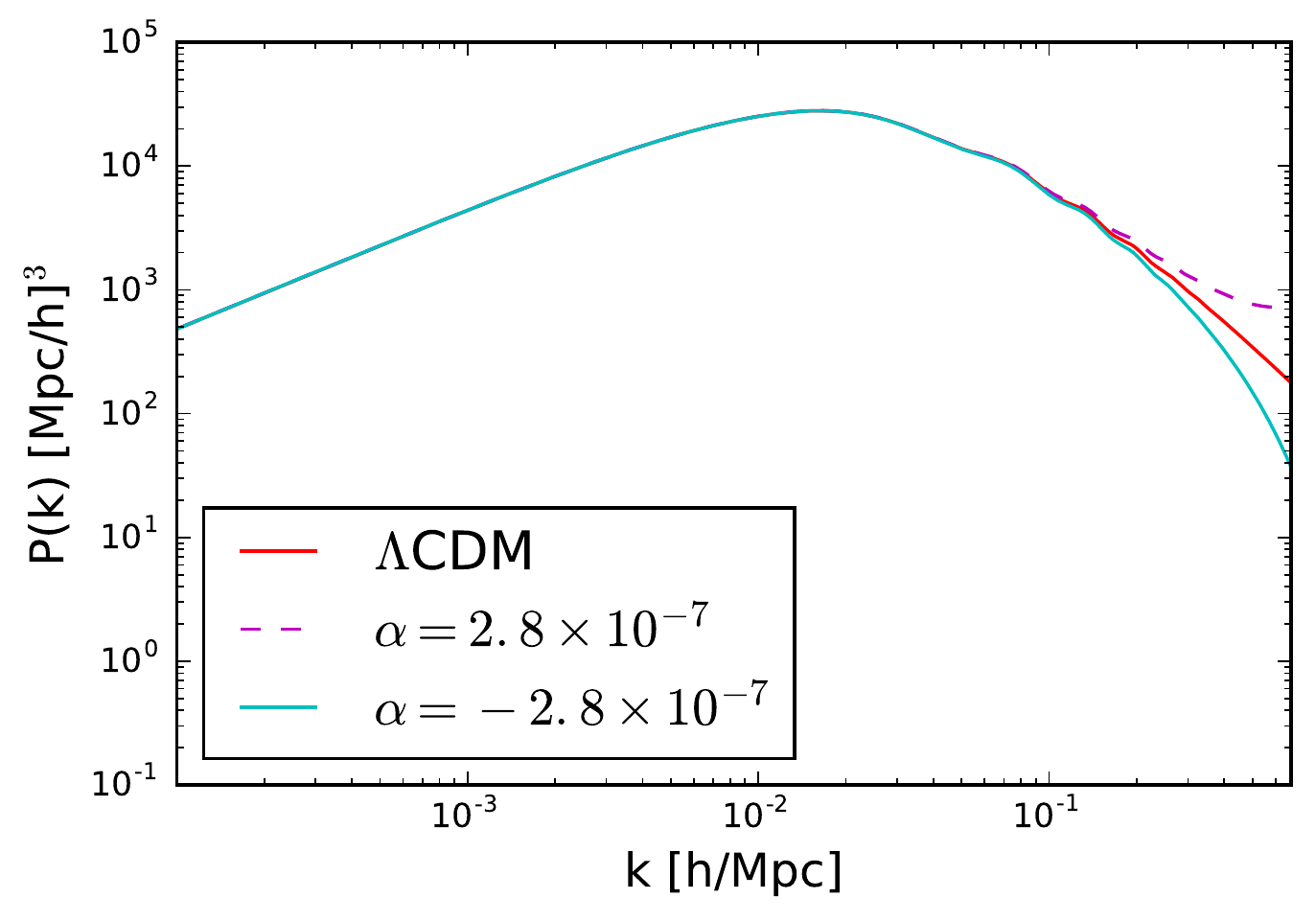}
        }
\subfigure[]{          \label{fig4b}
           \includegraphics[width=0.42\textwidth]{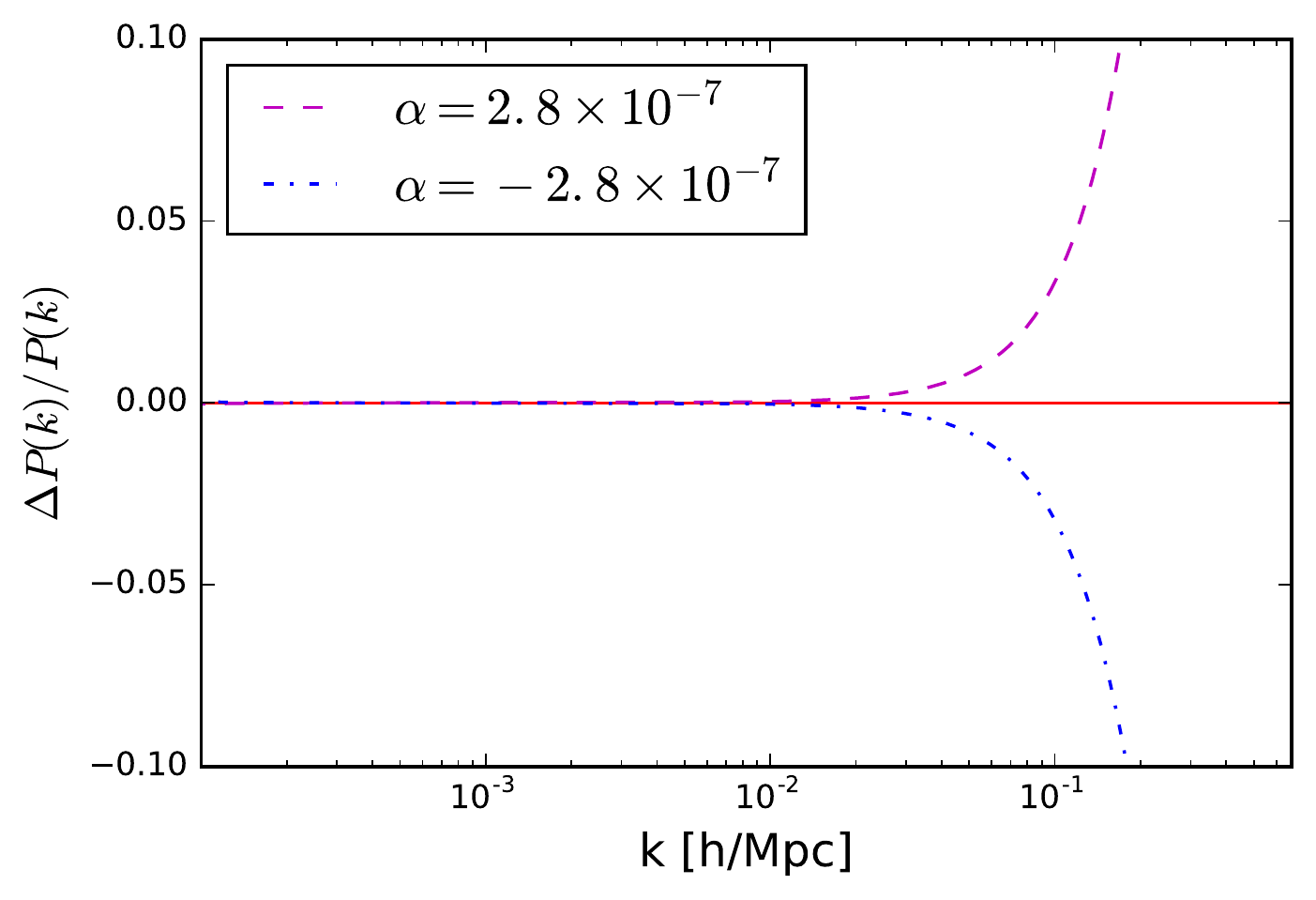}
        }
\end{center}
\caption{ (a) Theoretical prediction for the matter power spectrum in the presence of the pseudo nonminimal interactions in CDM and relativistic relics for different values of $\alpha$, and the $\Lambda$CDM model. (b) Same as in (a), but quantifying the deviations with respect to $\Lambda$CDM model. For drawing the graphs, we have chosen the best fit mean values from Table \ref{tab:results} for the baseline parameters. }
\end{figure}

Figures \ref{fig4a} and \ref{fig4b}  show the theoretical prediction for the matter power spectrum $P(k,z=0$), which depends on the evolution of the density perturbations $\delta_{\rm b}$, $\delta_{\rm cdm}$, and $\delta_{\nu}$, where the later two carry the effects of $\alpha$ corrections. As expected, varying $\alpha$ values, with order of magnitude $10^{-7}$, do not cause significant deviations at larger scales (where the linear effects are predominant) since for $k < k_{\rm nr}$ \footnote{$k_{\rm nr}$ is the free-streaming scale where the neutrinos become nonrelativistic. Here, we have $k_{\rm nr} = 0.003$ h/Mpc.} the matter power spectrum of a $\Lambda$CDM model with massive neutrinos is the same as that of a massless model (relativistic relics). Basically, the effects on the scales $k < k_{\rm nr}$ may be due to $w_{\rm cdm}$ and $n_s$. But, here, the corrections via the scale-independent EMSG do not change these parameters significantly, so significant changes on the linear scale are not expected.
On the other hand, at smaller scales and when $k \geq k_{\rm nr}$, we observe significant deviations even for very small $\alpha$ corrections. It is important to note that we have considered one massive neutrino with a fixed mass scale at $0.06$ eV in the present analysis. It is well known that neutrino masses produce a smooth suppression of the matter power spectrum on small scales. Thus, the suppression that we notice here is due to the combined effect of the massive neutrino at $0.06$ eV plus negative $\alpha$ value corrections on $\delta_{\rm cdm}$. For $k > k_{\rm nr}$, relativistic species fluctuations remain smaller than CDM and baryon perturbations because of their low growth rates and are too small to backreact on $P(k)$. For positive $\alpha$ value corrections, which originally can induce an excess of density of CDM particles, we can have an increment on the amplitude of the matter spectrum on a small scale. These interesting features of our model could be useful while dealing with the small-scale issues of the standard $\Lambda$CDM model. However, it may be noted that the small-scale predictions on the matter power spectrum need further attention since the predictions here are based on linear perturbations of the background. Nonetheless, the $\alpha $ corrections in our model can indeed change the behavior of the matter power spectrum significantly at the smaller scales. 

\section{Final remarks}
We have introduced a scale-independent energy-momentum squared gravity model, which allows different gravitational couplings for different types of sources and new redshift dependencies of the energy densities of them without the need of invoking nonminimal interactions of them with some other sources. We have then introduced an extension of the $\Lambda$CDM model, where photons and baryons couple to the spacetime as in GR, while the dark sector components, namely, CDM and relativistic relics, couple to the spacetime in accordance with the scale-independent EMSG, and $\Lambda$ is a bare cosmological constant (conventional vacuum energy is assumed to be null). This scenario induces pseudo nonminimal interactions on CDM and relativistic relics leading to modification on their dynamics at both the background and perturbative levels. We have found that the observational constraints on the model parameters from the CMB and BAO data which suggest that the induced corrections due to the scale-independent EMSG are not statistically significant; viz., the dimensionless constant $\alpha$ of the model is constrained to the order $10^{-7}$, and thereby, the deviations from the $\Lambda$CDM model are minimal. For instance, the deviations on CMB TT are around 0.001 \%. On the other hand, even in this case, the model leads to a considerable deviation in the matter power spectrum from the standard $\Lambda$CDM at smaller scales, where the nonlinear effects are predominant, as may be seen in Fig. \ref{fig4a}. This interesting feature of our model could have some implications on the small scale issues of the standard $\Lambda$CDM model and deserves further attention. We also have briefly discussed further possible applications/implications of the model for such small values as well as somewhat larger values of $\alpha$ in Sec. \ref{sec:obsresults}.

We would finally like to present a couple of examples, giving insight that our model would have many useful and interesting applications/implications depending on its dimensionless constant $\alpha$ beyond what we have found from somewhat conservative analysis of the model in the present work. To do so, rewriting the continuity equation \eqref{noncons} in the form $\dot \rho_i+3H \left(1+w_{{\rm eff},i}\right)\rho_i=0$ for a source described by an EoS parameter equal to $w_i$, we define a corresponding effective EoS parameter as
\begin{align}
\label{modcont}
w_{{\rm eff},i}=w_i+\frac{\alpha_i (1-w_i^2)}{4\alpha_i w_i+\sqrt{3w^2+1}},
\end{align}
which implies that the source $i$ with $w_i$ contributes to Einstein's field equations of GR like a source with an EoS parameter equal to $w_{{\rm eff},i}$. According to this, for instance, we have 
\begin{equation}
\label{eqn:effde}
w_{{\rm eff},\nu}=\frac{1}{3}+\frac{4\alpha_{\nu}\sqrt{3}}{6\alpha_{\nu}\sqrt{3}+9},
\end{equation}
for the relativistic species with EoS parameter $w_{\nu}=\frac{1}{3}$. The corresponding effective energy density and pressure, as can be seen from \eqref{eq:rhoprime} and \eqref{eq:presprime}, read $\rho_{{\rm eff},\nu}=(1+\frac{2\alpha}{\sqrt{3}}) \rho_{\nu}$ and $p_{{\rm eff},\nu}=(\frac{1}{3}+\frac{2\alpha}{\sqrt{3}})\rho_{\nu}$ and $\rho_{{\rm eff},\nu}>0$ provided that $\alpha>-\frac{3}{2\sqrt{3}}$. The relativistic species can then mimic a DE source when $\alpha_{\nu}\sim-\frac{1}{\sqrt{3}}$, such that in this case we have $\rho_{{\rm eff},\nu}>0$ and $w_{{\rm eff},\nu}={\rm constant}\sim-1$, and may account for the late time acceleration of the Universe like a $w$CDM type model. Moreover, provided that the cosmological model contains a cosmological constant as well, these relativistic species together with a cosmological constant would effectively contribute to Einstein's field equations of GR like a dynamical DE, the EoS parameter of which could be written as
\begin{equation}
w_{{\rm eff},\nu,\Lambda}=\frac{\left(\frac{1}{3}+\frac{2\alpha_{\nu}}{\sqrt{3}}\right)\rho_{\nu}-\Lambda}{\left(1+\frac{2\alpha_{\nu}}{\sqrt{3}}\right)\rho_{\nu}+\Lambda},
\end{equation}
where $\rho_{\nu}=\rho_{{\nu},0}\,(1+z)^{4+\frac{4\alpha}{2\alpha+\sqrt{3}}}$. It could then range between the constant given in \eqref{eqn:effde} depending on $\alpha$ (when the relativistic species are dominant) and minus unity (when the cosmological constant is dominant). We note that this EoS parameter is reminiscent of a canonical scalar field as $w_{{\rm eff},\nu,\Lambda}\sim \frac{\rho_{\nu}-\Lambda}{\rho_{\nu}+\Lambda}$ for $ \alpha_{\nu} \gg 0$ and, in general, of the one that has been obtained by introducing a noncanonical scalar field \cite{Mukhanov:2005bu}, which has, for instance, been considered for unifying dark matter and DE (see Ref. \cite{Mishra:2018tki} and references therein). Such interesting features of the scale-independent EMSG model lead us to stress that this model is quite rich and promising to further investigate its possible  applications/implications in cosmology.

\begin{acknowledgements}
 \"{O}.A. acknowledges the support by the Science Academy in scheme of the Distinguished Young Scientist Award  (BAGEP).  N.K. acknowledges the postdoctoral research support she is receiving from the Istanbul Technical University. S.K. gratefully acknowledges the support from SERB-DST Project No. EMR/2016/000258 and the warm hospitality and research facilities provided by the Inter-University Centre for Astronomy and Astrophysics, India, where a part of this work was carried out.
\end{acknowledgements}


\begin{thebibliography}{99}

\bibitem{Komatsu:2010fb} 
  E.~Komatsu {\it et al.} (WMAP Collaboration), Seven-Year Wilkinson Microwave Anisotropy Probe (WMAP) observations: Cosmological interpretation, Astrophys.\ J.\ Suppl.\ Ser.  \textbf{192}, 18 (2011), [arXiv:1001.4538].

\bibitem{Aubourg:2014yra} \'E.~Aubourg {\it et al.}, Cosmological implications of baryon acoustic oscillation measurements, Phys.\ Rev.\ D \textbf{92}, 123516 (2015), [arXiv:1411.1074].

\bibitem{Planck2015} P. A. R. Ade {\it et al.} (Planck collaboration), Planck 2015 results. XIII. Cosmological parameters, Astron. Astrophys. \textbf{594}, A13 (2016), [arXiv:1502.01589].

\bibitem{Weinberg:1988cp} S.~Weinberg, The cosmological constant problem, Rev. Mod. Phys. \textbf{ 61}, 1 (1989).

\bibitem{Peebles:2002gy} P.~J.~E.~Peebles and B.~Ratra, The Cosmological constant and dark energy, Rev.\ Mod.\ Phys. {\bf 75}, 559 (2003), [astro-ph/0207347].

\bibitem{Padmanabhan:2002ji}
  T.~Padmanabhan, Cosmological constant: The weight of the vacuum, Phys.\ Rep.\ \textbf{380}, 235 (2003), [hep-th/0212290].
 
\bibitem{Bullock:2017xww}
  J.~S.~Bullock and M.~Boylan-Kolchin, Small-Scale Challenges to the $\Lambda$CDM Paradigm, Annu.\ Rev.\ Astron.\ Astrophys. \textbf{55}, 343 (2017), [arXiv:1707.04256].
  
\bibitem{Freedman:2017yms}
W.~L.~Freedman, Cosmology at a crossroads, Nat.\ Astron.\  {\bf 1}, 0121 (2017), [arXiv:1706.02739].
  
\bibitem{tension01} A. Bhattacharyya, U. Alam, K. L. Pandey, S. Das and S. Pal, Are $H_0$ and $\sigma_8$ tensions generic to present cosmological data?, [arXiv:1805.04716].

\bibitem{tension02} M. Raveri and W. Hu, Concordance and discordance in Cosmology, [arXiv:1806.04649].

\bibitem{tension03} E. Di Valentino, Crack in the cosmological paradigm, Nat. Astron. \textbf{1}, 569 (2017), [arXiv:1709.04046].

\bibitem{Zhao:2017cud}
  G.~B.~Zhao {\it et al.}, Dynamical dark energy in light of the latest observations, Nature (London) \textbf{1}, 627 (2017), [arXiv:1701.08165].
  
  \bibitem{Copeland:2006wr}
E.~J.~Copeland, M.~Sami and S.~Tsujikawa, Dynamics of dark energy,  Int.\ J.\ Mod.\ Phys.\ D \textbf{15}, 1753 (2006), [hep-th/0603057].
  
\bibitem{Caldwell:2009ix}
  R.~R.~Caldwell and M.~Kamionkowski, The physics of cosmic acceleration, Annu.\ Rev.\ Nucl.\ Part.\ Sci.\  \textbf{59}, 397 (2009), [arXiv:0903.0866].
  
\bibitem{Clifton:2011jh}
T.~Clifton, P.~G.~Ferreira, A.~Padilla and C.~Skordis, Modified gravity and cosmology, Phys.\ Rep.\  \textbf{513}, 1 (2012), [arXiv:1106.2476].

\bibitem{DeFelice:2010aj} 
  A.~De Felice and S.~Tsujikawa, $f(R)$ theories, Living Rev. Relativity \textbf{13}, 3 (2010), [arXiv:1002.4928].
  
\bibitem{Capozziello:2011et} 
  S.~Capozziello and M.~De Laurentis, Extended theories of gravity, Phys. Rep. \textbf{509}, 167 (2011), [arXiv:1108.6266].

\bibitem{Nojiri:2017ncd} 
  S.~Nojiri, S.~D.~Odintsov, and V.~K.~Oikonomou, Modified gravity theories on a nutshell: Inflation, bounce and late-time evolution, Phys.\ Rep. \textbf{692}, 1 (2017), [arXiv:1705.11098].

\bibitem{Nojiri:2010wj} 
  S.~Nojiri and S.~D.~Odintsov, Unified cosmic history in modified gravity: From $F(R)$ theory to Lorentz non-invariant models, Phys. Rep. \textbf{505}, 59 (2011), [arXiv:1011.0544].
 \bibitem{Hui:2009kc} L.~Hui, A.~Nicolis and C.~Stubbs, Equivalence Principle Implications of Modified Gravity Models, Phys.\ Rev.\ D {\bf 80}, 104002 (2009)[arXiv:0905.2966].

 \bibitem{peebles10} J. A. Keselman, A. Nusser, and P. J. E. Peebles, Cosmology with equivalence principle breaking in the dark sector, Phys.\ Rev.\ D \textbf{81}, 063521 (2010), [arXiv:0912.4177].
 
\bibitem{Mohapi:2015gua} N.~Mohapi, A.~Hees and J.~Larena, Test of the equivalence principle in the dark sector on galactic scales, J. Cosmol. Astropart. Phys. {\bf 03}, (2016) 032 [arXiv:1510.06198].
  
\bibitem{DM_DE01} B. Wang, E. Abdalla, F. Atrio-Barandela, and D. Pavon, Dark matter and dark energy interactions: Theoretical challenges, cosmological implications and observational signatures, Rep. Prog. Phys. \textbf{79}, 096901 (2016), [arXiv:1603.08299].

\bibitem{DM_DE02} S. Kumar and R. C. Nunes, Echo of interactions in the dark sector, Phys. Rev. D \textbf{96}, 103511 (2017), [arXiv:1702.02143].
 
\bibitem{DM_DE03} W. Yang, S. Pan, E. Di Valentino, R. C. Nunes, S. Vagnozzi and D. F. Mota, [arXiv:1805.08252].

\bibitem{DM_nu01} R. J. Wilkinson, C. Boehm, and J. Lesgourgues, Constraining dark matter-neutrino interactions using the CMB and large-scale structure,  J. Cosmol. Astropart. Phys. \textbf{05}, (2014) 01, [arXiv:1401.7597].
 
\bibitem{DM_nu02} P. Serra, F. Zalamea, A. Cooray, G. Mangano and A. Melchiorri, Constraints on neutrino-dark matter interactions from cosmic microwave background and large scale structure data, Phys. Rev. D \textbf{81}, 043507 (2010), [arXiv:0911.4411].

\bibitem{DM_B01} C. Dvorkin, K. Blum and M. Kamionkowski, Constraining dark matter-baryon scattering with linear cosmology, Phys. Rev. D \textbf{89}, 023519 (2014), [arXiv:1311.2937].

\bibitem{DM_B02} Y. Bai, J. Salvado, and B. A. Stefanek, Cosmological constraints on the gravitational interactions of matter and dark matter,  J. Cosmol. Astropart. Phys. \textbf{10}, (2015) 029, [arXiv:1505.04789].

\bibitem{DM_B03} K. K. Boddy and V. Gluscevic, First Cosmological constraint on the effective theory of dark matter-proton interactions, [arXiv:1801.08609].

\bibitem{DM_gamma01} J. Stadler and C. Boehm, CMB constraints on $\gamma$-CDM interactions revisited, [arXiv:1802.06589].

\bibitem{DM_gamma02} S.~Kumar, R.~C.~Nunes and S.~K.~Yadav, Cosmological bounds on dark matter-photon coupling, [arXiv:1803.10229].

\bibitem{DM_DR01} F. Y. C. Racine, R. de Putter, A. Raccanelli, and K. Sigurdson, Constraints on large-scale dark acoustic oscillations from cosmology, Phys. Rev. D \textbf{89}, 063517 (2014), [arXiv:1310.3278].

\bibitem{DM_DR02} M. A. B. Abad, M. Schmaltz, J. Lesgourgues and T. Brinckmann, Interacting dark sector and precision cosmology,  J. Cosmol. Astropart. Phys. \textbf{01}, (2018) 008, [arXiv:1708.09406].

\bibitem{Katirci:2014sti} N.~Kat{\i}rc{\i} and M.~Kavuk, $f(R,T_{\mu\nu}T^{\mu\nu})$ gravity and Cardassian-like expansion as one of its consequences, Eur.\ Phys.\ J.\ Plus \textbf{129}, 163 (2014), [arXiv:1302.4300].

\bibitem{Harko:2010mv} T.~Harko and F.~S.~N.~Lobo, $f(R, \mathcal{L}_{\rm m}$) gravity, Eur.\ Phys.\ J.\ C \textbf{70}, 373 (2010), [arXiv:1008.4193].

 \bibitem{Harko:2011kv} 
  T.~Harko, F.~S.~N.~Lobo, S.~Nojiri and S.~D.~Odintsov, $f(R,T)$ gravity, Phys.\ Rev.\ D \textbf{84}, 024020 (2011), [arXiv:1104.2669].

\bibitem{Akarsu:2017ohj} \" O.~Akarsu, N.~Kat{\i}rc{\i}, and S.~Kumar, Cosmic acceleration in a dust only Universe via energy-momentum powered gravity, Phys.\ Rev.\ D \textbf{97}, 024011 (2018), [arXiv:1709.02367].

\bibitem{Akarsu:2018zxl} \" O.~Akarsu, J.~D.~Barrow, S. \c C{\i}k{\i}nto\u glu, K.~Y.~Ek\c si and N.~Kat{\i}rc{\i}, Constraint on energy-momentum squared gravity from neutron stars and its cosmological implications, Phys.\ Rev.\ D \textbf{97}, 124017 (2018), [arXiv:1802.02093].

\bibitem{Board:2017ign} C.~V.~R.~Board and J.~D.~Barrow, Cosmological models in energy-momentum-squared gravity, Phys.\ Rev.\ D \textbf{96}, 123517 (2017), [arXiv:1709.09501].


\bibitem{Roshan:2016mbt} M.~Roshan and F.~Shojai, Energy-momentum squared gravity, Phys.\ Rev.\ D \textbf{94}, 044002 (2016), [arXiv:1607.06049].


\bibitem{Nari:2018aqs}
  N.~Nari and M.~Roshan, Compact stars in energy-momentum squared gravity, Phys. Rev. D \textbf{98}, 024031 (2018), [arXiv:1802.02399].


\bibitem{Ashtekar:2006wn}   
 A.~Ashtekar, T.~Pawlowski and P.~Singh, Quantum nature of the big bang: Improved dynamics, Phys.\ Rev.\ D \textbf{74}, 084003 (2006), [gr-qc/0607039].

\bibitem{Ashtekar:2011ni} 
A.~Ashtekar and P.~Singh, Loop quantum cosmology: A status report, Classical  Quantum Gravity \textbf{28}, 213001 (2011),[arXiv:1108.0893].

\bibitem{Brax:2003fv} 
  P.~Brax and C.~van de Bruck, Cosmology and brane worlds: A review, Classical Quantum Gravity \textbf{20}, R201 (2003), [hep-th/0303095].
 
 \bibitem{Harko:2014aja} T.~Harko, F.~S.~N.~Lobo, G.~Otalora and E.~N.~Saridakis, $f(T,\mathcal{T})$ gravity and cosmology, J. Cosmol. Astropart. Phys. 12 (2014) 021, [arXiv:1405.0519 [gr-qc]].
 
\bibitem{Damour90} T. Damour, G. W. Gibbons, and C. Gundlach, Dark matter, time-varying G, and a dilaton field, Phys. Rev. Lett. \textbf{64}, 123 (1990).
  
\bibitem{Wetterich:1994bg} C.~Wetterich, The Cosmon model for an asymptotically vanishing time dependent cosmological 'constant', Astron.\ Astrophys.\ {\bf 301} 321 (1995), [hep-th/9408025].
  
\bibitem{amendola2000} L. Amendola, Phys. Rev. D \textbf{62}, 043511 (2000), [astro-ph/9908023].
\bibitem{amendola2002} L. Amendola and D. Tocchini-Valentini, Phys. Rev. D 66, 043528 (2002), [astro-ph/0111535].


\bibitem{Farrar:2003uw} G.~R.~Farrar and P.~J.~E.~Peebles, Interacting dark matter and dark energy,  Astrophys.\ J.\  {\bf 604} 1 (2004), [astro-ph/0307316].


\bibitem{GarciaBellido:1992de} J.~Garcia-Bellido, Dark matter with variable masses, Int.\ J.\ Mod.\ Phys.\ D {\bf 02} 85,  (1993), [hep-ph/9205216].

  
\bibitem{DM_nu03} E. Di Valentino, C. Boehm, E. Hivon, and F. R. Bouchet, Reducing the $H_0$ and $\sigma_8$ tensions with Dark Matter-neutrino interactions, Phys. Rev. D \textbf{97}, 043513 (2018), [arXiv:1710.02559].

\bibitem{DM_nu04} A. O. D. Campo, C. Boehm, S. P. Ruiz, and S. Pascoli, Dark matter-neutrino interactions through the lens of their cosmological implications, Phys. Rev. D \textbf{97}, 075039  (2018), [arXiv:1711.05283].

\bibitem{Neff01} R. C. Nunes and A. Bonilla, Probing the properties of relic neutrinos using the cosmic microwave background, the Hubble Space Telescope and galaxy clusters, Mon. Not. R. Astron. Soc.\textbf{473}, 4404  (2018), [arXiv:1710.10264].

\bibitem{Neff02} M. Gerbino and M. Lattanzi, Status of neutrino properties and future prospects-Cosmological and astrophysical constraints, Front. Phys. \textbf{5}, 70 (2017), [arXiv:1712.07109].

\bibitem{Lovelock:1971yv} D.~Lovelock, The Einstein tensor and its generalizations, J.\ Math.\ Phys.
\textbf{12}, 498 (1971).

\bibitem{Lovelock:1972vz} D.~Lovelock, The four-dimensionality of space and the Einstein tensor, J.\ Math.\ Phys. \textbf{13}, 874 (1972).


\bibitem{Bull:2015stt} P.~Bull {\it et al.}, Beyond $\Lambda$CDM: Problems, solutions, and the road ahead, Phys. Dark Univ. \textbf{12}, 56 (2016), [arXiv:1512.05356].
 
 \bibitem{Straumann} N.~Straumann, \textit{General Relativity: With Applications to Astrophysics} (Springer-Verlag, Berlin, 2004).


\bibitem{Bertolami:2008ab} O.~Bertolami, F.~S.~N.~Lobo, and J.~Paramos, Non-minimum coupling of perfect fluids to curvature, Phys.\ Rev.\ D \textbf{78}, 064036 (2008), [arXiv:0806.4434].

\bibitem{Faraoni:2009rk} V.~Faraoni, The Lagrangian description of perfect fluids and modified gravity with an extra force, Phys.\ Rev.\ D \textbf{80}, 124040 (2009), [arXiv:0912.1249].
  
\bibitem{Uzan:2010pm} 
  J.~P.~Uzan, Varying constants, gravitation and cosmology, Living Rev.\ Relativity, \textbf{14} 2 (2011), [arXiv:1009.5514].

\bibitem{Ma_Bertschinger} C. P. Ma and E. Bertschinger, Cosmological perturbation theory in the synchronous and conformal Newtonian gauges, Astrophys. J. \textbf{455}, 7 (1995), [astro-ph/9506072].

\bibitem{class} D. Blas, J. Lesgourgues, and T. Tram, The Cosmic Linear Anisotropy Solving System (CLASS) II: Approximation schemes,  J. Cosmol. Astropart. Phys. \textbf{07}, 034 (2011), [arXiv:1104.2933].

\bibitem{xi01} J. Lesgourgues and S. Pastor, Cosmological implications of a relic neutrino asymmetry, Phys. Rev. D, \textbf{60}, 103521 (1999), [hep-ph/9904411].

\bibitem{xi02}E. Castorina, U. Fran\c ca, M. Lattanzi, J. Lesgourgues, G.
Mangano, A. Melchiorri, and S. Pastor, Cosmological lepton asymmetry with a nonzero mixing angle $\theta_{13}$, Phys. Rev. D \textbf{86}, 023517 (2012), [arXiv:1204.2510].

\bibitem{Salas} P. F. de Salas, M. Lattanzi, G. Mangano, G. Miele, S. Pastor and O. Pisanti, Bounds on very low reheating scenarios after Planck, Phys. Rev. D \textbf{92}, 123534 (2015), [arXiv:1511.00672].

\bibitem{Kawasaki} M. Kawasaki, K. Kohri, and N. Sugiyama, MeV-scale reheating temperature and thermalization of neutrino background, Phys. Rev. D \textbf{62}, 023506 (2000), [astro-ph/0002127].

\bibitem{Mangano} G. Mangano, G. Miele, S. Pastor, T. Pinto, O. Pisanti, and P. D. Serpico, Effects of non-standard neutrino-electron interactions on relic neutrino decoupling, Nucl. Phys. B\textbf{756}, 100 (2006), [hep-ph/0607267].

\bibitem{p13} P. A. R. Ade {\it et al.} (Planck Collaboration), Planck 2013 results. XVI. Cosmological parameters, Astron. Astrophys. \textbf{571}, A16 (2014), [arXiv:1303.5076].

\bibitem{bao1} F. Beutler, C. Blake, M. Colless, D. H. Jones, L. Staveley-Smith, L. Campbell, Q. Parker, W. Saunders, and F. Watson, The 6dF Galaxy Survey: Baryon Acoustic Oscillations and the Local Hubble Constant, Mon. Not. R. Astron. Soc. \textbf{416}, 3017 (2011), [arXiv:1106.3366].

\bibitem{bao2}  A. J. Ross, L. Samushia, C. Howlett, W. J. Percival, A. Burden, and M. Manera, The clustering of the SDSS DR7 Main Galaxy Sample I: A 4 per cent distance measure at $z=0.15$, Mon. Not. R. Astron. Soc. \textbf{449}, 835 (2015), [arXiv:1409.3242].

\bibitem{bao3} L. Anderson {\it et al.}, The clustering of galaxies in the SDSS-III Baryon Oscillation Spectroscopic Survey: Baryon acoustic oscillations in the Data Release 10 and 11 galaxy samples, Mon. Not. R. Astron. Soc. \textbf{441}, 24 (2014), [arXiv:1312.4877].

\bibitem{bao4} A. Font-Ribera {\it et al.}, Quasar-Lyman $\alpha$ Forest cross-correlation from BOSS DR11 : Baryon acoustic oscillations,  J. Cosmol. Astropart. Phys. \textbf{5}, 27 (2014), [arXiv:1311.1767].

\bibitem{coupled04} R. C. Nunes, S. Pan, and E. N. Saridakis, New constraints on interacting dark energy from cosmic chronometers, Phys. Rev. D \textbf{94}, 023508 (2016), [arXiv:1605.01712].

\bibitem{monte} B. Audren, J. Lesgourgues, K. Benabed and S. Prunet, Conservative constraints on early cosmology: An illustration of the Monte Python cosmological parameter inference code,  J. Cosmol. Astropart. Phys. \textbf{02} (2013) 001, [arXiv:1210.7183].

\bibitem{Gelman} A. Gelman and D. Rubin, Inference from iterative simulation using multiple sequences, Stat. Sci., \textbf{7}, 457 (1992).

\bibitem{antonygetdist} \href{https://github.com/cmbant/getdist}{https://github.com/cmbant/getdist}

\bibitem{Abad} M. A. B. Abad, G. M. Tavares and M. Schmaltz, Non-Abelian dark matter and dark radiation, Phys. Rev. D \textbf{92}, 023531 (2015), [arXiv:1505.03542].

\bibitem{Weinberg02} S. Weinberg, Goldstone Bosons as Fractional Cosmic Neutrinos, Phys. Rev. Lett., \textbf{110}, 241301 (2013), [arXiv:1305.1971]. 

\bibitem{Abazajian} K. N. Abazajian {\it et al.}, Light sterile neutrinos: A white paper, [arXiv:1204.5379].

\bibitem{Mukhanov:2005bu} V.~F.~Mukhanov and A.~Vikman, Enhancing the tensor-to-scalar ratio in simple inflation, J. Cosmol. Astropart. Phys. \textbf{02} (2006) 004, [astro-ph/0512066].
 
 \bibitem{Mishra:2018tki} S.~S.~Mishra and V.~Sahni, Unifying dark matter and dark energy with non-canonical scalars, [arXiv:1803.09767].  

  

  
\end{thebibliography}
\end{document}